\documentstyle[12pt]{article}

\advance \textheight by \topskip
\makeatletter
\def\eqnarray{\stepcounter{equation}\let\@currentlabel=\theequation
\global\@eqnswtrue
\global\@eqcnt\z@\tabskip\@centering\let\\=\@eqncr
$$\halign to \displaywidth\bgroup\@eqnsel\hskip\@centering
  $\displaystyle\tabskip\z@{##}$&\global\@eqcnt\@ne
  \hfil$\displaystyle{\hbox{}##\hbox{}}$\hfil
  &\global\@eqcnt\tw@ $\displaystyle\tabskip\z@
  {##}$\hfil\tabskip\@centering&\llap{##}\tabskip\z@\cr}
\@addtoreset{equation}{section}
  \def\theequation{\thesection.\arabic{equation}}
\makeatother
\def\im{\mathop{\rm Im}\nolimits}
\def\ker{\mathop{\rm Ker}\nolimits}
\def\rank{\mathop{\rm rank}\nolimits}
\mathchardef\by="0202
\def\bm#1{\mbox{\boldmath $#1$}}

\title{Dirac versus reduced phase space quantization\\for
systems admitting no gauge conditions}

\author{Mikhail S. Plyushchay${}^{a}$\thanks{On leave from the Institute for
High Energy Physics, Protvino, Russia; E--mail:
mikhail@cc.unizar.es}$\ $, Alexander V.
Razumov${}^{b}$\thanks{E--mail: razumov@mx.ihep.su}\\ [1ex]{\it
${}^{a}$Departamento de F\'{\i}sica Te\'orica, Facultad de
Ciencias}\\ {\it Universidad de Zaragoza, 50009 Zaragoza,
Spain}\\ [0.5ex]{\it ${}^{b}$Institute for High Energy Physics,
142284 Protvino,} \\ {\it Moscow Region, Russia}\\ [0.5ex]PACS:
03.65.Ca, 02.40.-k}

\date{To appear in {\bf Inter. J. Mod. Phys. A}}

\begin{document}

\maketitle

\begin{abstract}
The constrained Hamiltonian systems admitting no gauge conditions are
considered. The methods to deal with such systems are discussed and
developed. As a concrete application, the relationship between the Dirac
and reduced phase space quantizations is investigated for spin models
belonging to the class of systems under consideration.  It is traced out
that the two quantization methods may give similar, or essentially
different physical results, and, moreover,
a class of constrained systems, which can be
quantized only by the Dirac method, is discussed.
A possible interpretation of the gauge degrees of freedom is
given.

\end{abstract}
\newpage

\section{Introduction}

Nowadays there are two main methods to quantize the Hamiltonian systems
with first class constraints: the Dirac quantization \cite{Dir} and the
reduced phase space quantization \cite{Fad69}, while the path integral
method \cite{FPo67,Fad69} and the BRST quantization, being the most
popular method for the covariant quantization of gauge--invariant systems
\cite{brst}, are based on and proceed from them \cite{Fad69,mar,sun}. The
basic idea of the Dirac method consists in imposing quantum mechanically
the first class constraints as operator conditions on the states for
singling out the physical ones \cite{Dir}. The reduced phase space
quantization first identifies the physical degrees of freedom at the
classical level by the factorization of the constraint surface with
respect to the action of the gauge group, generated by the constraints.
Then the resulting Hamiltonian system is quantized as a usual
unconstrained system \cite{Fad69}.  Naturally, the problem of the
relationship of these two methods arises. It was discussed in different
contexts in literature (see \cite{ash}--\cite{ord}, and references
therein), and it was demonstrated for the concrete examples that these two
methods give, in general, different results. The corresponding quantum
systems may differ, e.g., in their energy level structure, or in the
eigenvectors of their Hamiltonians \cite{rom,sch}. Nevertheless, there is
an opinion that the differences between the two quantization methods can
be traced out to a choice of factor ordering in the construction of
various physical operators.

In the present paper we investigate the relationship of the two methods of
quantization for the special class of Hamiltonian systems with first class
constraints corresponding to different physical models of spinning
particles.  The specific general property of the considered examples of
constrained systems is the following: their constraints generate SO(2)
transformations and, hence, corresponding gauge orbits topologically are
one-spheres $S^{1}$. This fact implies that these systems {\it do not
admit gauge conditions}.  Nevertheless, for the first three systems we can
construct the corresponding reduced phase spaces. We shall demonstrate
that the two methods of quantization give coinciding physical results only
for the first of these systems, but they give, in general, {\it
essentially different physical results} for two other systems.

Since in the present paper we consider the constrained systems which do
not admit gauge conditions, we should use a general geometrical approach to
the Dirac--Bergmann theory of the constrained systems for the construction
of the reduced phase spaces.  This approach is reviewed in Section 2 of
our paper.  Note that from the point of view of this approach, it is not
necessary to use the notion of the weak equality of the functions on the
phase space, and this notion is not used in our paper at all.

The first concrete example, we consider, is the plane spin model, which is
a subsystem of the (3+1)--dimensional models of massless particles with
arbitrary helicity \cite{ply2}, and of the (2+1)--dimensional relativistic
models of fractional spin particles \cite{ply3}. This model is related
also to the simplest nonrelativistic model of the anyon \cite{wil}.  The
classical and quantum treatment of the model having a cylinder
$S^{1}\times {\bf R}$ as the initial phase space, is performed in Section
3.  Here we trace out an interesting analogy in interpretation of the
situation with nonexistence of a global gauge condition for this simple
constrained system with the situation taking place for the non-Abelian
gauge theories \cite{sin}.

The second model, which is considered in Section 4, is the rotator spin
model, being a subsystem of the (3+1)--dimensional model of the
relativistic massive particle with integer spin \cite{ply4}.  The
nontriviality of the model reveals itself in the topology of the
constraint surface diffeomorphic to the group manifold of the Lie group
SO(3).

Section 5 is devoted to the consideration of the top spin model, which, in
turn, is a subsystem in the non--Grassmannian model of the relativistic
massive particle with arbitrary integer or half--integer spin \cite{ply5}.
This system is also topologically nontrivial and has the cotangent bundle
$T^{*}{\rm SO}(3)$ as the initial phase space.

In section 6 we summarize the results, present conclusions on the
relationship of the two methods of quantization and formulate a possible
interpretation of the sense of gauge degrees of freedom for constrained
systems. Here we also discuss a broad class of pseudoclassical models
\cite{bri}--\cite{cor}, i.e., models containing Grassmann variables,
whose general feature consists in the presence of the constraints,
nonlinear in Grassmann variables and having no nontrivial projections on
the unit of Grassmann algebra. For such systems it is not possible to
construct the reduced phase space in principle, and they can be quantized
{\it only by the Dirac method}.

Appendixes A and B are devoted to the presentation of the method of
dependent coordinates used in the paper. Actually this method is, in a
sense, a reformulation of the Dirac bracket method, but from our point of
view it has, in comparison with the former, greater mathematical
definiteness, and turns out to be more convenient for our purposes.

Appendix C collects the facts about three dimensional rotation group, which
are needed in the main text.

Everywhere in the text repeated indices imply the corresponding summation.

\section{Geometrical Background of the Dirac--Bergmann Theory}

As it has been noted in the Introduction, the systems we consider in the
present paper do not admit gauge conditions. To construct the
reduced phase space for them we should use a general geometrical approach
(see \cite{AbM78}, and references therein).  This approach together with
its relation to the usual Dirac--Bergmann formalism is briefly described
below.

Let $(M, \omega^M)$ be an $m$--dimensional symplectic manifold, and $N$ be
an $n$--dimensional submanifold of $M$. Let $\iota: N \to M$ be the
inclusion mapping. The two--form $\omega^N = \iota^* \omega^M$ is
obviously closed. Suppose that $\omega^N$ has a constant rank, then the
set
\[
{\cal E}_{\omega^N} = \{ x \in TN \mid i(x) \omega^N = 0\}
\]
is a distribution on $N$, called the characteristic distribution of
$\omega^N$. As $\omega^N$ is closed, ${\cal E}_{\omega^N}$ is integrable
and by the Frobenius theorem defines a foliation ${\cal F}_{\omega^N}$ on
$N$. Identifying all points of a leaf we get the quotient space $P =
N/{\cal F}_{\omega^N}$. Assume that $P$ is a manifold with the canonical
projection $\pi: N \to P$ being a submersion.
In other words, we suppose
that $\pi: N \to P$ is a fibered manifold.
In this case it can be shown that
there exists a closed nondegenerate two--form $\omega^P$, such that
\begin{equation}
\pi^* \omega^P = \omega^N. \label{2}
\end{equation}
Thus, the manifold $P$ has a natural structure of a symplectic manifold.

In the Dirac--Bergmann theory the submanifold $N$ is a manifold defined by
equations
\[
\psi_a = 0, \qquad a = 1,\ldots, m-n.
\]
It means that
\[
N = \{ p\in M \mid \psi_a(p) = 0,\, a = 1,\ldots,m-n\},
\]
where $\psi_a$ are differentiable functions on $M$, such that the mapping
$\psi: M \to {\bf R}^{m-n}$, defined by $p \to (\psi_1(p), \ldots,
\psi_{m-n}(p))$ is of rank $m-n$ for every $p \in N$.

Let $x \in T_p N$, then
\[
d\psi_a(\iota_*(x)) = 0, \qquad a=1,\ldots,m-n.
\]
On the other hand, suppose that $y \in T_{\iota(p)} M$, $p\in N$, and
\begin{equation}
d\psi_a(y) = 0, \qquad a = 1,\ldots, m-n,
\label{5}
\end{equation}
then there exists a unique vector $x \in T_p N$, such that $y =
\iota_*(x)$.

Recall that the Hamiltonian vector field $X_f$ on $M$, corresponding to
the function $f$ on $M$, is defined by
\begin{equation}
i(X_f) \omega^M = df, \label{6}
\end{equation}
and for the Poisson bracket of two functions, $f$ and $g$, on $M$ we have
the expression
\begin{equation}
\{f, g\} = -\omega^M(X_f, X_g) = X_f(g) = -X_g(f). \label{7}
\end{equation}

Introduce the matrix valued function $\Delta = \|\Delta_{ab}\|$,
where
\[
\Delta_{ab} = \{\psi_a, \psi_b\},
\]
and define
\[
\widetilde \Delta_{ab} = \iota^* \Delta_{ab}.
\]
Suppose, that the matrix $\widetilde \Delta$ is degenerate at some point
$p \in N$. Hence we can find a nontrivial set of real numbers $c^a$,
$a = 1,\ldots, m-n$, such that
\[
c^a \widetilde \Delta_{ab}(p) = 0.
\]
{}From Eq.~(\ref{7}) we get
\[
c^a X_{a\iota(p)} (\psi_b) = d\psi_{b\iota(p)}(c^a
X_a) = 0,
\]
where we have denoted $X_{\psi_a}$ through $X_a$. Thus, there
exists a vector $x \in T_p N$, such that
\begin{equation}
\iota_*(x) = c^a X_{a\iota(p)}. \label{11}
\end{equation}
Let us show that the vector $x$ belongs to ${\cal E}_{\omega^N}$. Indeed,
for any $x' \in T_n N$ we have
\[
[i(x)\omega^N](x') = \omega^N(x,x') = \omega^M(\iota_*(x), \iota_*(x')).
\]
Using Eq.~(\ref{11}), we get
\[
[i(x) \omega^N](x') = c^a \omega^M (X_a, \iota_*(x')) = c^a
d\psi_a (\iota_*(x')) = 0.
\]
That was to be demonstrated.

On the other hand, suppose that $x \in {\cal E}_{\omega^N}$. If $x \in
T_p N$, then for any $x' \in T_p N$ we have
\[
0 = [i(x) \omega^N](x') = \omega^M(\iota_*(x), \iota_*(x')) =
[i(\iota_*(x))\omega^M](\iota_*(x')).
\]
Note that if a one--form $\mu \in T^*_\iota(p) M$ is such that
\[
\mu(\iota_*(x)) = 0
\]
for all $x \in T_p N$, then
\[
\mu = c^a d\psi_{a\iota(p)}
\]
for some constants $c^a$, $a = 1,\ldots, m-n$. Hence, we have
\[
\iota_*(x) = c^a X_{a\iota(p)},
\]
and
\[
0 = d\psi_b(\iota_*(x)) = c^a X_{a\iota(p)} (\psi_b) =
c^a \widetilde \Delta_{ab}(p).
\]
Thus, we see that the elements of ${\cal E}_{\omega^N}$ and the null
vectors of the matrices $\widetilde \Delta(p)$, $p \in N$, are in
one--to--one correspondence, and the two--form $\omega^N$ is of constant
rank if and only if the rank of $\widetilde \Delta(p)$ is independent of
$p$.

Recall that in the Dirac--Bergmann theory the functions $\psi_a$ are
constraints, the submanifold $N$ is called the constraint surface, and the
symplectic manifold $P$ is called the reduced phase space.

A function $f$ on $M$ is said to be of first class, if
\[
\{f, \psi_a \} = f^b_a \psi_b
\]
for some functions $f^b_a$. A constraint being a function of first class,
is called the first class constraint. Suppose that the matrix $\widetilde
\Delta(p)$ is of rank $k$ at any point $p \in M$. Let the
indices $\alpha$, $\beta$, $\gamma$ take values from $1$ to
$k$, while the index $\mu$ takes values from $k+1$ to $m-n$.
Without a loss of generality, we can consider, in fact locally,
that the submatrix $\|\Delta_{\alpha \beta}(p)\|$ is
nondegenerate at any point $p \in M$. In this case we can
introduce the functions $\Delta^{\alpha \beta}$ satisfying the
relation \[ \Delta^{\alpha \gamma}(p) \Delta_{\gamma \beta}(p)
= \delta^\alpha_\beta \] for all $p \in N$. It is clear that
the functions \begin{eqnarray*} &&\phi_\alpha = \psi_\alpha, \\
&&\phi_\mu = \psi_\mu - \{\psi_\mu, \psi_\alpha\} \Delta^{\alpha \beta}
\psi_\beta
\end{eqnarray*}
form an equivalent set of constraints defining the constraint surface $N$.

The constraints $\phi_\mu$ are first class constraints. It can be easily
shown that the Hamiltonian vector fields $X_{\phi_\mu}$ are tangent to
$N$, and these fields restricted to $N$ form a basis of the
characteristic distribution ${\cal E}_{\omega^N}$.  The main
property of the constraints $\phi_\alpha$, which certainly
follows from their definition, is that the matrix formed by the
Poisson brackets of $\phi_\alpha$ is nondegenerate at any point
of the constraint surface. The constraints $\phi_\alpha$ are
called second class constraints.

If the matrix $\widetilde \Delta(p)$ is of zero rank at any point $p \in
N$, we have only first class constraints. In this case
\begin{equation}
\{\psi_a, \psi_b\} = \psi_c f^c{}_{ab}
\label{18a}
\end{equation}
for some functions $f^c{}_{ab}$ on $M$. If the matrix $\widetilde
\Delta(p)$ is nondegenerate at any point of $N$ we deal with second class
constraints. In this case the corresponding characteristic distribution is
trivial, and the reduced phase space coincides with the manifold $N$. A
convenient method to work with second class constraints is the Dirac
bracket method, or the method of dependent coordinates (see Appendix B).
Note that in this case it is convenient to use upper indices for the
constraints.

Let us discuss now the procedure of constructing the reduced phase space
for the case of first class constraints, according to the Dirac--Bergmann
theory.

Below we identify the tangent space $T_p N$ with the set of vectors from
$T_{\iota(p)} M$, satisfying Eq.~(\ref{5}), and consider the submanifold
$N$ simply as a subset of $M$. If the matrix $\widetilde \Delta(p)$ is of
rank zero at any point $p \in N$, then the Hamiltonian vector fields $X_a$
belong to the characteristic distribution ${\cal E}_{\omega^N}$, and
generate it. In the case when the functions $f^c{}_{ab}$ from
Eq.~(\ref{18a}) are constant on $M$, we can try to find a group of
canonical transformations $G$ such that the functions $\psi_a$ be its
generators. This group is called the gauge group, generated by the
constraints $\psi_a$. If such a group exists, the leaves of the foliation,
defined by the characteristic distribution, are its orbits.

Suppose that the fibered manifold $\pi: N \to P$ has a global section $s:
P \to N$. The differentiable mapping $s$ by definition satisfies the relation
\begin{equation}
\pi \circ s = \mbox{id}_P. \label{19}
\end{equation}
It is clear that the image $s(P)$ of the
section
$s$ can be considered as
a submanifold of $N$ that is diffeomorphic to $P$. Furthermore, from
Eqs.~(\ref{2}) and (\ref{19}) it follows that
\[
s^* \omega^N = \omega^P.
\]
Hence,
\[
s^* \omega^{s(P)} = \omega^P,
\]
where $\omega^{s(P)}$ is the restriction of the two--form $\omega^N$ to
the submanifold $s(P)$, and $s$ is considered as a mapping from $P$ to
$s(P)$. Thus the submanifold $s(P)$ is a symplectic manifold that is
symplectomorphic to $P$, and we can identify $P$ with $s(P)$.
Choosing different global sections of the fibered manifold $\pi: N \to P$,
we come to different realizations of the reduced phase space as a
submanifold of $N$. Note that $N$ is a submanifold of the symplectic
manifold $M$, hence we actually have realizations of the reduced phase
space as a submanifold of $M$.

Assume now that the reduced phase space, realized as a submanifold of $M$,
is defined by the equations
\[
\psi_a = 0, \qquad \chi^a = 0, \qquad a = 1,\ldots, m-n.
\]
The functions $\chi^a$ are called gauges, or gauge conditions. They
can be considered as additional constraints. Denote the full set of the
constraints by $\psi^\alpha$, $\alpha = 1,\ldots,2(m-n)$:
\[
\psi^a = \psi_a, \qquad \psi^{a+(m-n)} = \chi^a.
\]
As $P$ is a symplectic submanifold of $M$, the matrix $\widetilde
\Xi(p)$ with the matrix elements $\widetilde \Xi^{\alpha\beta} =
\iota^* \{\psi^\alpha, \psi^\beta\}(p)$ is nondegenerate at any point $p
\in P$. From this it follows that the matrix
\begin{equation}
\widetilde \Lambda_a{}^b(p) = \iota^*\{\psi_a, \chi^b\}(p) \label{23}
\end{equation}
must be of rank $m-n$ at any point of $p \in P$.

Note that a
section $s$ allowing
to identify the reduced phase space
$P$ with a symplectic submanifold $s(P)$ of the initial symplectic
manifold $M$ may not exist.
In this case we say that the system under consideration does not admit
gauge conditions.  In such a situation we can always consider a set of
local sections $s_i: U_i \to N$, covering $P$. For any $i$ the set
$s_i(U_i)$ is symplectomorphic to the open symplectic submanifold $U_i$.
The total reduced phase space can be considered as the result of natural
patching of submanifolds $s_i(U_i)$.  Actually it is more convenient to
construct the reduced phase space by the direct geometrical construction
described above. It is the method we use in the present paper.

Note also that even if a global section of the fibered manifold $\pi: N
\to P$ exists we
cannot, in general, consider $s(P)$ as a submanifold of $M$, defined by
equations.
We shall encounter such a situation in the next section.

\section{Plane Spin Model}

The first model we are going to consider is the plane spin model
\cite{ply2,ply3}. This is a very simple model, but nevertheless here we
come across peculiarities relevant for a more general case. Therefore, we
give for this model a more detailed consideration than it deserves from a
first sight. The initial phase space of the model is a cotangent bundle
$T^* S^1$ of the one--dimensional sphere $S^1$, that is a cylinder
$S^{1}\times {\bf R}$. It can be described locally by an angular variable $0
\le \varphi < 2\pi$ and the conjugate momentum $S\in {\bf R}$. Here the
angle variable cannot be considered as a global coordinate. The symplectic
two--form $\omega$ in terms of the local variables $\varphi$, $S$ has the
form
\[
\omega = dS \wedge d \varphi.
\]
Thus, we have locally
\[
\{\varphi, S\} = 1.
\]
Actually, any $2\pi$--periodical function of the variable $\varphi$ that
is considered as a variable, taking values in ${\bf R}$, can be considered
as a function on the phase space, i.e., as an observable, and any
observable is connected with the corresponding $2\pi$--periodical
function. Therefore, we can introduce the functions
\begin{equation}
q_1 = \cos \varphi, \qquad q_2 = \sin \varphi \label{2.2a}
\end{equation}
as the functions on the phase space of the system. For these functions we
have
\begin{eqnarray}
&\{q_1, q_2\}= 0,& \nonumber \\[-2ex]
\label{2.3} \\[-2ex]
&\{q_1, S\} = - q_2, \qquad \{q_2, S\} = q_1.& \nonumber
\end{eqnarray}
The functions $q_1$ and $q_2$ are not independent, and satisfy
the relation
\[
q_1^2 + q_2^2 = 1.
\]
Note, that any function on the phase space can be considered as a function
of dependent coordinates $q_1$, $q_2$ and $S$.  These coordinates will be
taken below as the quantities, forming a restricted set of observables
whose quantum analogs have the commutators which are in the direct
correspondence with their Poisson brackets.

We come to the plane spin model by introducing the `spin' constraint
\begin{equation}
\psi = S - \theta = 0, \label{2.9}
\end{equation}
where $\theta$ is an arbitrary real constant.

Consider now the Dirac quantization of the system. To this end let us take
as the Hilbert space the space of complex $2\pi$--periodical functions of
the variable $\varphi$ with the scalar product
\begin{equation}
(\Phi_1,\Phi_2) = \frac{1}{2\pi} \int_0^{2\pi}
\overline{\Phi_1(\varphi)}\Phi_2 (\varphi)\, d\varphi. \label{2.5}
\end{equation}
The operators $\hat q_1$ and $\hat q_2$, corresponding to the functions
$q_1$ and $q_2$, are the operators of multiplication by the functions
$\cos \varphi$ and $\sin\varphi$, respectively, i.e.,
\[
\hat q_1 \Phi = q_1 \Phi, \qquad \hat q_2 \Phi = q_2 \Phi.
\]
The operator $\hat S$ is defined by
\[
\hat S \Phi = \left(-i \frac{d}{d \varphi} + c\right)\Phi,
\]
where $c$ is an arbitrary real constant. It is clear that the operators
$\hat q_1$, $\hat q_2$ and $\hat S$ are Hermitian operators with respect
to the scalar product, defined by Eq.~(\ref{2.5}).

One can easily show that the relations
\begin{eqnarray*}
&[\hat q_1, \hat q_2] = 0, \nonumber \\[-2ex]
\label{2.8} \\[-2ex]
&[\hat q_1, \hat S] = -i \hat q_2, \qquad [\hat q_2, \hat S] = i \hat
q_1& \nonumber
\end{eqnarray*}
are valid for any value of $c$ in agreement with Eq.~(\ref{2.3}).

The quantum analog of the constraint (\ref{2.9}) gives the equation for
the physical state wave functions:
\begin{equation}
(\hat S - \theta)\Phi_{phys} = 0. \label{2.10}
\end{equation}
Decomposing the function $\Phi_{phys}(\varphi)$ over the orthonormal basis,
formed by
the functions $e^{ik\varphi}$:
\[
\Phi_{phys}(\varphi)=\sum_{k \in {\bf Z}} \psi_k e^{ik\varphi},
\]
we find that equation (\ref{2.10}) has a nontrivial solution only when
\[
c = \theta + n,
\]
where $n$ is some fixed integer, $n \in {\bf Z}.$ In this case the
corresponding physical normalized wave function is
\begin{equation}
\Phi_{phys}(\varphi) = e^{in\varphi}. \label{2.13}
\end{equation}
Here the only physical operator \cite{sun},
i.e. an operator commuting with the
quantum constraint $\hat \psi$, is $\hat S$, and it is
reduced to the constant $\theta$ on the physical subspace
(\ref{2.13}).

Now we come back to the classical theory in order to construct the reduced
phase space of the model. Let us show that for the surface, defined by
Eq.~(\ref{2.9}), there is no `good' gauge condition, but, nevertheless, the
reduced phase space of the system can be constructed. Indeed, it is clear
that the one--parameter group of transformations, generated by
the constraint $\psi$, consists of the rotations of the phase
space. This group acts transitively on the constraint surface,
and we have only one gauge orbit, which is the constraint
surface itself. The gauge conditions must single out one point
of an orbit. In our case we have to define only one gauge
condition, let us denote it by $\chi$. The function $\chi$ must
be such that the equations \[ \psi = 0, \qquad \chi = 0 \]
determine a set, consisting of only one point, and in this point we
should have
\[
\{\psi, \chi\} \ne 0
\]
(see Section 2). Recall that any function on the phase space of
the system can be considered as a
function of the variables $\varphi$ and $S$, which is
$2\pi$--periodical with respect to $\varphi$.  Thus, we require
the $2\pi$--periodical function $\chi(\varphi, S)|_{S=\theta}$
turns into zero at only one point $\varphi = \varphi_0$ from
the interval $0 \le \varphi < 2\pi$. Moreover, we should have
\[ \{\psi, \chi\}(\varphi_0, \theta) = - \left.\frac{\partial
\chi(\varphi, \theta)}{\partial \varphi}\right|_{\varphi =
\varphi_0} \ne 0.  \] It is clear that such a function does not
exist.  Nevertheless, we have here the reduced phase space that
consists of only one point. Therefore, the reduced space
quantization is trivial: physical operator $\hat S$ takes here
constant value $\theta$ in correspondence with the results
obtained by the Dirac quantization method.  When the described planar spin
model is a subsystem of some other system, the reduction means simply that
the cylinder $T^{*}S^{1}$ is factorized into a point, where $S = \theta$,
and that wave functions do not depend on the variable $\varphi$.

Let us point out one interesting analogy in interpretation of the
situation with nonexistence of a global gauge condition. Here the
condition of $2\pi$--periodicity can be considered as a `boundary'
condition. If for a moment we forget about it, we can take as a gauge
function any monotonic function $\chi(\varphi, S)$, $\chi \in {\bf R}$,
such that $\chi(\varphi_{0}, \theta)=0$ at some point $\varphi =
\varphi_{0}$, and, in particular, we can choose the function
$\chi(\varphi, S) = \varphi$. The `boundary' condition excludes all such
global gauge conditions. In this sense the situation is similar to the
situation in the non--Abelian gauge theories where without taking into
account the boundary conditions for the fields it is also possible to find
global gauge conditions, whereas the account of those leads, in the end, to
the nonexistence of global gauge conditions (see, e.g., \cite{sin}).

Now we are going to describe the system using dependent coordinates (see
Appendix B). To this end let us take two--dimensional vectors $\bm q$ and
$\bm p$, where $\bm p$ is defined as
\[
p_1 = -S\sin \varphi, \qquad p_2 = S \cos \varphi.
\]
Let $\epsilon_{ij}$ be a skew symmetric tensor, normalized by the
condition
$
\epsilon_{12} = 1.
$
For any two dimensional vectors $\bm a$ and $\bm b$ we use the
notation
\[
\bm a \times \bm b = \epsilon_{ij} a_i b_j.
\]
It is clear that
\[
S = \epsilon_{ij} q_i p_j = \bm q \times \bm p.
\]

The variables $\bm q$ and $\bm p$ satisfy the relations
\[
\bm q^2 = 1, \qquad \bm q \bm p = 0,
\]
and can be considered as dependent coordinates in the phase space.
For nonzero Poisson brackets we have the following expressions:
\begin{equation}
\{p_i, p_j\} = p_i q_j - q_i p_j, \qquad \{q_i, p_j\} = \delta_{ij} - q_i
q_j. \label{2.21}
\end{equation}
Hence, the Hamiltonian vector field, corresponding to constraint
(\ref{2.9}) has, according to Eq.~(\ref{b.*}), the form
\[
X_\psi = - \bm q \times \frac{\partial}{\partial \bm q} - \bm p \times
\frac{\partial}{\partial \bm p}.
\]
This vector field generates the characteristic distribution of the
reduction of the symplectic two--form to the constraint surface, given by
Eq.~(\ref{2.9}). The corresponding leaves are the orbits of the
one--parameter group of canonical transformations, generated by the
function $\psi$. To find them we can use Eq.~(\ref{b.25a}), that leads to
\[
\bm q(\tau) = \bm q \cos \tau + S^{-1} \bm p \sin \tau, \qquad
\bm p(\tau) = \bm p \cos \tau - S \bm q \sin \tau.
\]
It is clear that in accordance with the above consideration the constraint
surface consists of only one leaf, and the reduced phase space is a point.

Concluding this section, let us write the corresponding Lagrangian for the
described Hamiltonian system:
\begin{equation}
L = - \theta \dot {\bm q} \times \bm q - \lambda (\bm q^2 - 1), \label{2.25}
\end{equation}
where $\lambda$ is a Lagrange multiplier. It is not difficult to show that
this Lagrangian leads to the Hamiltonian system with two second class
constraints
\[
\psi_1 = \bm q \bm p = 0, \qquad \psi_2 = \frac{1}{2}(\bm q^2 - 1) = 0,
\]
and one first class constraint
\[
\psi = \bm q\times \bm p - \theta = 0.
\]
Using the notion of the Dirac bracket (see Appendix B), it is easy to get
convinced that the Poisson brackets for the variables $q_i$ and $p_i$,
restricted to the submanifold, defined by the constraints $\psi_1$ and
$\psi_2$, coincide with those, given by Eq.~(\ref{2.21}). It is convenient
further to use the same letters for coordinates and their restrictions to
the submanifold under consideration. The exact meaning of the coordinates
is always clear from the context.

Using parameterization (\ref{2.2a}), we come from Lagrangian (\ref{2.25})
to the Lagrangian
\begin{equation}
L=\theta\dot{\varphi}.
\label{f11}
\end{equation}
Note that Lagrangian (\ref{f11}) is present as a part of a complete
Lagrangian in the nonrelativistic model of the anyon \cite{wil}.

Concluding this section, stress once more that for the considered simplest
system the both quantization methods give coinciding results.

\section{Rotator Spin Model}

In this section we consider the rotator spin model \cite{ply4}. The
initial phase space of the system is described by a spin three--vector
$\bm S$ and a unit vector $\bm q$,
\[
\bm q^2=1,
\]
being orthogonal one to the other,
\[
\bm q \bm S = 0.
\]

The variables $q_i$ and $S_i$, $i=1,2,3$, can be considered as dependent
coordinates in the phase space of the system. The Poisson brackets for
these coordinates are
\[
\{q_i, q_j\} = 0, \quad \{S_i, S_j\} = \epsilon_{ijk}S_k, \quad \{S_i,
q_j\} = \epsilon_{ijk} q_k.
\]
Introducing the vector
\[
\bm p = \bm S \times \bm q,
\]
we get for its components the following Poisson brackets:
\[
\{p_i, p_j\} = p_i q_j - q_i p_j, \qquad \{q_i, p_j\} = \delta_{ij} - q_i
q_j,
\]
that is of the same form as Eq.~(\ref{2.21}), and
\[
\{S_i, p_j\} = \epsilon_{ijk} p_k.
\]
Note that the vector $\bm p$ satisfies the relation
\[
\bm q \bm p = 0.
\]
Actually, the variables $q_i$ and $p_i$ can be considered as another set
of dependent coordinates in the phase space.

Using the expressions for the Poisson brackets of the dependent variables,
we find the following expression for the symplectic two--form:
\begin{equation}
\omega = d p_i \wedge d q_i = d(\epsilon_{ijk} S_j q_k) \wedge d q_i.
\label{dpdq}
\end{equation}
Note that the phase space under consideration is the internal phase space
of the one--mode relativistic string model \cite{PrR83}.

Introducing the spherical angles $\varphi$, $\vartheta$ ($0 \le \varphi <
2\pi,\, 0 \le \vartheta \le \pi)$ and the corresponding momenta
$p_\varphi, p_\vartheta \in {\bf R}$, we can write the following
parameterization for the vectors $\bm q$ and $\bm p$:
\begin{equation}
\bm q = \left(\begin{array}{c}
 \cos \varphi \sin \vartheta \\
 \sin \varphi \sin \vartheta \\
 \cos \vartheta
\end{array}\right), \qquad
\bm p = \left(\begin{array}{c}
 \cos \varphi \cos \vartheta p_\vartheta - \displaystyle \frac{\sin
  \varphi}{\sin \vartheta} p_\varphi \\[1.7ex]
 \sin \varphi \cos \vartheta p_\vartheta + \displaystyle \frac{\cos
  \varphi}{\sin \vartheta} p_\varphi \\ [1.5ex]
 - \sin \vartheta p_\vartheta
\end{array}\right), \label{3.10}
\end{equation}
and for the symplectic two--form we get the expression
\[
\omega = d p_\vartheta \wedge d \vartheta + d p_\varphi \wedge d\varphi.
\]
{}From this relation we conclude that the initial phase space of the system
is symplectomorphic to the cotangent bundle $T^*S^2$ of the
two--dimensional sphere $S^2$, furnished with the canonical symplectic
structure.

The rotator spin model is obtained from the initial phase space by
imposing the constraint
\begin{equation}
\psi = \frac{1}{2}(\bm S^2 - \rho^2) = 0, \qquad \rho > 0, \label{r4}
\end{equation}
fixing the spin of the system. This can be obtained also starting from the
Lagrangian
\[
L = -\rho \sqrt{\dot{\bm q}{}^2} - \lambda ({\bm q}^{2} - 1),
\]
where $\lambda$ is a Lagrange multiplier. Indeed, the Hamiltonian
description of the system with this Lagrangian results in two second class
constraints
\[
\psi_1 = \bm q \bm p = 0, \qquad \psi_2 = \frac{1}{2} (\bm q^2 - 1) = 0,
\]
and a first class constraint (\ref{r4}) with $\bm S = \bm q \times \bm p$.
The reduction to the surface, defined by the second class constraints
$\psi_1$ and $\psi_2$, leads to the Poisson brackets which coincide with
those given above.

Using the Dirac method, we quantize the model in the following way.
The state space is a space of the square integrable functions on the
two--dimensional sphere. The scalar product is
\[
(\Phi_1, \Phi_2) = \int_{S^2} \overline{\Phi_1(\varphi, \vartheta)}
\Phi_2(\varphi, \vartheta) \sin \vartheta d\vartheta d\varphi.
\]
Parameterization (\ref{3.10}) allows us to use as the operator $\hat {\bm
S}$ the usual orbital angular momentum operator expressed via spherical
angles \cite{Mes61}. The wave functions as the functions on a sphere are
decomposable over the complete set of the spherical harmonics:
\[
\Phi(\varphi, \vartheta) =
\sum_{l=0}^{\infty} \sum_{m=-l}^{l} \Phi_{lm} Y^l_m(\varphi, \vartheta),
\]
and, therefore, the quantum analog of the first class constraint (\ref{r4}),
\begin{equation}
(\hat {\bm S}{}^2 - \rho^2) \Phi_{phys} = 0, \label{r9}
\end{equation}
leads to the quantization condition for the constant $\rho$:
\begin{equation}
\rho^2 = n(n+1), \label{r10}
\end{equation}
where $n > 0$ is an integer.
Only in this case equation (\ref{r9}) has a nontrivial solution of the form
\[
\Phi_{phys}^n(\varphi, \vartheta) = \sum_{m=-n}^n \Phi_{nm} Y^n_m
(\varphi, \vartheta),
\]
i.e., with the choice of (\ref{r10}) we get the states with spin equal to $n$:
\[
\hat{\bm S}{}^2 \Phi_{phys}^n = n(n+1) \Phi_{phys}^n.
\]
Thus, we conclude that the Dirac quantization leads to the quantization
(\ref{r10}) of the parameter $\rho$ and, as a result, the quantum system
describes the states with integer spin $n$.

Let us turn now to the construction of the reduced phase space of the system.
Then we perform the reduced phase space quantization and reveal that this
method of quantization gives results physically different from those
obtained by the Dirac quantization method.

The constraint surface of the model can be considered as a set composed of
the points specified by two orthogonal normed three--vectors. Each pair of
such vectors can be supplemented by a unique third three--vector,
defined in such a way that we get an oriented orthonormal basis in three
dimensional vector space. It is well known that the set of all oriented
orthonormal bases in three dimensional space can be smoothly parameterized
by the elements of the Lie group SO(3) (see Appendix C). Thus, the
constraint surface in our case is diffeomorphic to the group manifold of
the Lie group SO(3).

As in the preceding Section, to find the one parameter group of canonical
transformations, generated by the constraint $\psi$, we can use
Eq.~(\ref{b.25a}). That gives the result
\[
\bm q(\tau) = \bm q \cos(S \tau) + \frac{1}{S}
\bm S \times \bm q \sin(S \tau), \qquad \bm S(\tau) = \bm S, \qquad S =
\sqrt{\bm S^2}.
\]
Hence, we see that the transformations we are looking for, are the
rotations about the direction, given by the spin vector. Thus, in the case
of a general position the orbits of the one--parameter group of
transformations under consideration are one dimensional spheres. Note,
that only the orbits, belonging to the constraint surface where $S = \rho
\ne 0$, are interesting to us. It is clear that an orbit is uniquely
specified by the direction of the spin three--vector $\bm S$ whose length
is fixed by the constraint $\psi$. As a result of our consideration, we
conclude that the reduced phase space of the rotator spin model is the
coset space SO(3)/SO(2), which is diffeomorphic to the two--dimensional
sphere $S^2$ \cite{War83}.  Due to the reasons discussed in the preceding
section there is no gauge condition in this case either. In fact, since
SO(3) is a nontrivial fiber bundle over $S^2$, we can neither find a
mapping from $S^2$ to SO(3) whose image would be diffeomorphic to the
reduced phase space. In other words, in this case the reduced phase space
cannot be considered as a submanifold of the constraint surface.

Our next goal is to write an expression for the symplectic two--form on
the reduced phase space. We shall give here two considerations of this
question, using independent and dependent coordinates.

Convenient independent coordinates in the initial phase space can be
introduced here with the help of the relations
\[
\bm S = S \left(\begin{array}{c}
 \sin\vartheta\cos\varphi \\
 \sin\vartheta\sin\varphi \\
 \cos\vartheta
\end{array}\right),\qquad
\bm q = \left(\begin{array}{c}
 \cos\gamma\cos\vartheta\cos\varphi-\sin\gamma\sin\varphi \\
 \cos\gamma\cos\vartheta\sin\varphi+\sin\gamma\cos\varphi \\
 -\cos\gamma\sin\vartheta
\end{array}\right),
\]
where $ 0 \le \vartheta \le \pi$, $0 \leq \varphi, \gamma< 2\pi$,
and $0 < S < \infty$. In terms of these coordinates the symplectic
two--form $\omega$ takes the form
\begin{equation}
\omega = d (S d\gamma + S \cos \vartheta d \varphi). \label{3.29}
\end{equation}
{}From this relation we get the following expressions for the nonzero
Poisson brackets
\begin{eqnarray*}
&\{\gamma, S\} = 1,& \\
&\{\varphi, \cos \vartheta\} = S^{-1}, \qquad \{\cos \vartheta,
\gamma\} = S^{-1} \cos \vartheta.&
\end{eqnarray*}
The restriction of the symplectic two--form $\omega$ to the constraint
surface, which we denote by the same letter, has the form
\begin{equation}
\omega = \rho d(\cos \vartheta d\varphi). \label{3.30}
\end{equation}

It is clear that the transformations of the group, generated by $\psi$,
act on the independent coordinates as follows:
\[
S(\tau) = S, \quad \vartheta(\tau) = \vartheta,\quad \varphi(\tau) =
\varphi, \quad \gamma(\tau) = \gamma + S \tau.
\]
Thus, we see that the variables $\vartheta$, $\varphi$ can be considered
as coordinates in the reduced phase space, and the symplectic two--form on
the reduced phase space is given by (\ref{3.30}).

The same results, but in a more transparent form, are obtained using
dependent coordinates. It is clear that we can consider the variables
$S_i$ as dependent coordinates in the reduced phase space, thus the
symplectic two--form on it may be expressed in terms of them. Note, to
this end, that the vectors $\bm q$, $\bm s = \bm S/S$ and $\bm q \times
\bm s$ form an orthonormal basis, and the differentials of the vectors
$\bm q$ and $\bm s$ satisfy the relations
\begin{eqnarray}
&\bm q d\bm q = 0, \qquad \bm s d\bm s = 0,& \label{3.31} \\
&\bm s d \bm q + \bm q d \bm s = 0.& \label{3.32}
\end{eqnarray}
Write the restriction of the symplectic two--form (\ref{dpdq}) to the
constraint surface as
\[
\omega = \rho (d \bm s \times \bm q) \wedge d \bm q + \rho (\bm s \times d
\bm q) \wedge d \bm q.
\]
{}From Eq.~(\ref{3.31}) it follows that we can write
\begin{eqnarray}
d \bm q &=& a \bm s + b (\bm q \times \bm s), \label{3.34} \\
d \bm s &=& f \bm q + g (\bm q \times \bm s), \label{3.35}
\end{eqnarray}
where $a$, $b$, $f$ and $g$ are some one--forms. From
Eq.~(\ref{3.34}) we get
\[
\bm s \times d \bm q = b \bm q,
\]
and, therefore,
\[
(\bm s \times d \bm q) \wedge d \bm q = 0.
\]
Further, from Eq.~(\ref{3.35}) we have
\begin{equation}
d \bm s \times \bm q = g \bm s.
\end{equation}
This equality implies
\[
(d \bm s \times \bm q) \wedge d \bm q = f \wedge g,
\]
where we have used Eq.~(\ref{3.32}). From Eq.~(\ref{3.35}) we have also
\[
(\bm s \times d \bm s) \wedge d \bm s = -2 f \wedge g,
\]
that gives finally
\begin{equation}
\omega = - \frac{1}{2\rho^2} (\bm S \times d \bm S) \wedge d \bm S.
\label{omega1}
\end{equation}
Thus, we see that the dependent coordinates $S^i$,
\begin{equation}
\bm S^2 = \rho^2,
\label{rho2}
\end{equation}
in the reduced phase space of the system provide a realization of the basis
of the Lie algebra so(3):
\begin{equation}
\{S_i, S_j\} = \epsilon_{ijk} S_k.
\label{sss}
\end{equation}

The quantization on the reduced phase space can be performed with the help
of the geometric quantization method proceeding from the classical relations
(\ref{omega1})--(\ref{sss}). This was done in detail earlier
(see, e.g., Ref.~\cite{pl6}), and we write here only the final results of
this procedure. The constant $\rho$ is quantized:
\begin{equation}
\rho = j, \qquad 0 < 2j \in {\bf Z}, \label{r22}
\end{equation}
i.e., it can take only integer or half-integer value,
and the Hermitian operators, corresponding to the components of the spin
vector, are realized in the form:
\begin{equation}
\hat S_1 = \frac{1-z^2}{2} \frac{d}{dz}+jz,\quad
\hat S_2 = i\left(\frac{1+z^2}{2}\frac{d}{dz} - jz\right),\quad
\hat S_{3} = z\frac{d}{dz} - j,
\label{r23}
\end{equation}
where
\[
z = e^{-i\varphi} \tan \vartheta/2,
\]
or, in terms of the dependent coordinates,
\[
z = \frac{S_1 - iS_2}{\rho + S_3}.
\]
Operators (\ref{r23}) act in the space of holomorphic functions $f(z)$
with the scalar product
\[
(f_1,f_2) = \frac{2j+1}{\pi} \int\int \frac{\overline{f_1(z)}f_2(z)}{(1 +
\vert z\vert^{2})^{2j+2}} d^{2}z,
\]
in which the functions
\[
\psi^{m}_{j}\propto z^{j+m},\qquad
m=-j,-j+1,...,j,
\]
form the set of eigenfunctions of the operator $\hat S_{3}$ with the
eigenvalues $s_{3}=m$.
These operators satisfy the relation
\[
\hat{\bm S}{}^2 = j(j+1)
\]
instead of classical relation (\ref{rho2}), and, therefore,
we have the $(2j+1)$--dimensional irreducible representation $D_j$ of the
Lie group SU(2).

Thus, we see that for the rotator spin model the reduced phase space
quantization method leads to the states with integer or half--integer
spin, depending on the choice of the quantized parameter $\rho$, and gives
in general the results physically different from the results obtained with
the help of the Dirac quantization method.  Let us stress once again here
that within the Dirac quantization method in this model the spin operator
$\hat{\bf S}$ has a nature of the orbital angular momentum operator, and
it is this nature that does not allow spin to take half-integer values
\cite{bied}.  In other words, in the Dirac quantization we quantize the
system with the phase space being the {\it cotangent bundle} to the {\it
two--dimensional sphere}, defined by the relation ${\bm q}^2 = 1$, while
in the reduced phase space quantization we start with the phase space
being a {\it two--dimensional sphere} of the radius coinciding with the
length of the vector ${\bm S}$. It is this difference between the phase
spaces that leads to the different results of quantizations.

\section{Top Spin Model}

In this section we consider the top spin model \cite{ply5}. The initial phase
space of the model is described by the spin three--vector $\bm S$, and by
three vectors $\bm e_i$, forming a right orthonormal basis in ${\bf
R}^3$:
\begin{equation}
\bm e_i \bm e_j = \delta_{ij}, \qquad \bm e_i \times \bm e_j =
\epsilon_{ijk} \bm e_k \label{5.1}
\end{equation}
(see Appendix C). Denote the components of the vectors $\bm e_i$ with
respect to the canonical basis of ${\bf R}^3$ by $E_{ij}$. The components
$S_i$ of the vector $\bm S$ and the quantities $E_{ij}$ form a set of
dependent coordinates in the phase space of the system. The corresponding
Poisson brackets are
\begin{eqnarray}
&\{E_{ij}, E_{kl}\} = 0,& \label{5.2} \\
&\{S_i, E_{jk}\} = \epsilon_{ikl}E_{jl}, \qquad \{S_i, S_j\} =
\epsilon_{ijk} S_k.& \label{5.3}
\end{eqnarray}
The vectors $\bm e_i$, satisfying Eq.~(\ref{5.1}) form a right orthonormal
basis in ${\bf R}^3$. The set of all such bases can be identified with the
three--dimensional rotation group (see Appendix C). Taking into account
Eqs.~(\ref{5.2}) and (\ref{5.3}) we conclude that the initial phase space
is actually the cotangent bundle $T^*{\rm SO(3)}$, represented as the
manifold ${\bf R}^3 \times {\rm SO(3)}$.
Using Eqs.~(\ref{5.2}) and (\ref{5.3}), one can get the
following expression for the symplectic two--form $\omega$ on the initial
phase space:
\begin{equation}
\omega = \frac{1}{2} d ((\bm S \times \bm e_l) d \bm e_l)) = \frac{1}{2}
d(\epsilon_{ijk} S_j E_{lk} d E_{li}). \label{5.4}
\end{equation}
It is useful to introduce the variables $J_i = \bm e_i \bm S = E_{ij} S_j$.
For these variables we have the following Poisson brackets:
\[
\{J_i, E_{jk}\} = - \epsilon_{ijl} E_{lk}, \qquad \{J_i, J_j\} =
-\epsilon_{ijk} J_k.
\]
Note, that we have the equality
\[
S_i S_i = J_i J_i.
\]

The phase space of the top spin model is obtained from the phase space,
described above, by introducing two first class constraints
\begin{eqnarray}
\psi &=& \frac{1}{2}(\bm S^2 - \rho^2) = 0,\label{sv1}\\
\chi &=& \bm S \bm e_3 - \kappa = 0,
\label{sv2}
\end{eqnarray}
where
\[
\rho > 0, \qquad |\kappa| < \rho.
\]

The described Hamiltonian system can be also obtained
starting from the Lagrangian
\begin{eqnarray}
L&=&-\theta\sqrt{{\dot{\bm q}}_1^2+{\dot{\bm q}}_2^2-({\dot{\bm
q}}_1 {\bm q}_2)^2-({\dot{\bm q}}_2 {\bm
q}_1)^2}-\frac{\kappa}{2}({\bm q}_1 {\dot{\bm
q}}_2-{\bm q}_2 {\dot{\bm q}}_1 )\nonumber\\
&{}&-\lambda_1 ({\bm q}_1^2 -1)-\lambda_2 ({\bm q}_2^2 -1)
-\lambda_3 {\bm q}_1 {\bm q}_2 , \nonumber
\end{eqnarray}
where
$\lambda_1$, $\lambda_2$ and $\lambda_3$ are Lagrange
multipliers and $\theta\neq 0$ is a real constant.
The Hamiltonian description of the system given by this
Lagrangian results in the set of six second class constraints
\begin{eqnarray}
&\psi_1={\bm q}_1^2 -1=0,\qquad \psi_2={\bm
q}_2^2 -1=0,\qquad \psi_3={\bm q}_1 {\bm q}_2 =0,&\nonumber\\
&\psi_4={\bm q}_1{\bm p}_1 =0,\qquad \psi_5={\bm q}_2{\bm
p}_2=0,\qquad \psi_6={\bm q}_1 {\bm p}_2 + {\bm q}_2 {\bm
p}_1=0,&\nonumber
\end{eqnarray}
and two  first class
constraints (\ref{sv1}) and (\ref{sv2}) with
$\rho^2=\theta^2+\kappa^2$, ${\bm S}={\bm q}_1\times{\bm p}_1 +
{\bm q}_2\times {\bm p}_2$, and ${\bm e}_3={\bm e}_1 \times
{\bm e}_2$, ${\bm e}_1={\bm q}_1$, ${\bm e}_2={\bm q}_2$.
The reduction to the surface of second class constraints
$\psi_1,\ldots,$ $\psi_6$ leads, in the end, to the
description of the system by dependent variables $E_{ij}$ and
$S_i$ having the Poisson brackets of the form (\ref{5.2}) and
(\ref{5.3}).

Consider now the Dirac quantization of the model. Recall that the matrix
$E$ can be identified with the corresponding rotation matrix (see Appendix
C). Let us parameterize the matrix $E$ by the Euler angles, $E = E(\alpha,
\beta, \gamma)$, and use the representation where the operators,
corresponding to these angles are diagonal. In this representation state
vectors are functions of the Euler angles, and the operators $\hat S_i$ and
$\hat J_i$ are realized as linear differential operators, acting on such
functions \cite{var}. The quantum analogs of the constraints $\psi$ and
$\chi$ turn into the equations for the physical states of the system:
\begin{eqnarray}
&(\hat{\bm S}{}^2 - \rho^2) \Phi_{phys} = 0,& \label{5.9} \\
&(\hat J_3 - \kappa) \Phi_{phys} = 0.& \label{5.10}
\end{eqnarray}

An arbitrary state vector can be decomposed over the set of the Wigner
functions, corresponding to either integer or half--integer spins
\cite{var}:
\begin{equation}
\Phi(\alpha, \beta, \gamma) = \phi_{jmk} D^j_{mk}(\alpha, \beta, \gamma),
\label{5.11}
\end{equation}
where $j = 0, 1, \ldots$, or $j = 1/2, 3/2, \ldots$, and $k, m = -j, -j+1,
\ldots, j$. The Wigner functions $D^j_{mk}$ have the properties
\[
\hat{\bm S}{}^2 D^j_{mk} = j(j+1) D^j_{mk}, \quad \hat S_3 D^j_{mk} = m
D^j_{mk}, \quad \hat J_3 D^j_{mk} = k D^j_{mk}.
\]
{}From the representation of the state vector (\ref{5.11}) we see that
Eqs.~(\ref{5.9}) and (\ref{5.10}) have nontrivial solutions only when
$\rho^2 = j(j+1)$, and $\kappa = k$, for some integer or half--integer
numbers $j$ and $k$, such that $-j \le k \le j$. In other words we get the
following quantization condition for the parameters of the model:
\[
\rho^{2} = j(j+1), \quad \kappa = k, \quad -j \le k \le j, \quad 0 < 2j \in
{\bf Z}.
\]
The corresponding physical state vectors have the form
\[
\Phi_{phys}(\alpha,\beta,\gamma) = \sum_{m=-j}^j \varphi_m
D^j_{mk}(\alpha, \beta, \gamma).
\]
Thus, we see that the Dirac quantization of the top spin model leads to
a pure integer or half--integer spin system.

Proceed now to the construction of the reduced phase space of the system.
As the constraints $\psi$ and $\chi$ have zero Poisson bracket, we can
consider them consecutively. Let us start with the constraint $\psi$.

{}From the expressions for the Poisson brackets (\ref{5.2}) and (\ref{5.3})
it follows that the group of gauge transformations, generated by the
constraint $\psi$, acts in the initial phase space variables as follows:
\begin{eqnarray}
&\bm e_i(\tau) = \bm e_i \cos (S \tau) + \frac{1}{S} (\bm S \times \bm e_i)
\sin (S \tau) + \frac{1}{S^2} \bm S (\bm S \bm e_i) (1 - \cos (S \tau)),&
\label{5.15} \\
&\bm S (\tau) = \bm S, \qquad S = \sqrt{\bm S^2}.& \label{5.16}
\end{eqnarray}
Comparing these relations with Eq.~(\ref{c.17}), we see that the
transformations under consideration have the sense of the rotation by the
angle $S\tau$ about the direction of the spin vector. In matrix notations
we can write
\begin{equation}
E(\tau) = E R^T(\bm s, S\tau), \qquad \bm s = \bm S/S. \label{5.17}
\end{equation}

Recall that the initial phase space of the system is diffeomorphic to
${\bf R}^3 \times {\rm SO(3)}$. Let us consider this space as a trivial
fibre bundle over ${\bf R}^3$ with the fibre SO(3). Gauge transformations
(\ref{5.15}), (\ref{5.16}) act in fibres of this bundle. It is
clear that the constraint surface, defined by the constraint $\psi$, is a
trivial fibre subbundle $S^2 \times {\rm SO(3)}$. As ${\rm SO(3)/SO(2)} =
S^2$, then after the reduction over the action of the gauge group we come
to the fibre bundle over $S^2$ with the fibre $S^2$. As it follows from
general theory of fibre bundles \cite{Hus66}, this fibre bundle is again
trivial. Thus the reduced phase space, obtained using only the constraint
$\psi$, is the direct product $S^2 \times S^2$. To find the expression for
the symplectic two--form on this reduced space we first introduce
convenient coordinates in the initial phase space.

Let ${\cal R}(\bm s, \bm n_3)$ be a rotation that transforms the vector
$\bm n_3$ to the vector directed along the vector $\bm s$. This rotation
is undefined for $S = 0$, but it is irrelevant for us, because we are
finally interested only in the constraint surface where $S = \rho \ne 0$.
We can choose ${\cal R}(\bm s, \bm n_3)$ in the form
\[
{\cal R}(\bm s, \bm n_3) = {\cal R}(\bm n_3, \varphi) {\cal R}(\bm n_2,
\vartheta),
\]
where $\varphi$ and $\vartheta$ are the polar angles of the vector $\bm S$.
We shall also use for the ${\cal R}(\bm s, \bm n_3)$ the notation ${\cal
R}(\varphi, \vartheta)$.
{}From the definition of ${\cal R}(\bm s, \bm n_3)$ we have
\begin{equation}
\bm S = S {\cal R}(\bm s, \bm n_3) \bm n_3. \label{5.19}
\end{equation}
Using now Eq.~(\ref{c.18a}), we can present the matrix $R(\bm s, S\tau)$
as
\[
R(\bm s, S \tau) = R(\bm s, \bm n_3) R(\bm n_3, S\tau) R^T(\bm s, \bm
n_3).
\]
Using this representation in Eq.~(\ref{5.17}), we get the following
transformation law for the matrix $F = E R(\bm s, \bm n_3)$:
\begin{equation}
F(\tau) = F R^T(\bm n_3, S\tau). \label{5.21}
\end{equation}
The matrix $F$ defines a new right basis $\{\bm f_i\}$, with
\begin{equation}
\bm f_i = F_{ij} \bm n_j = {\cal R}^{-1}(\bm s, \bm n_3) \bm e_i.
\label{5.21a}
\end{equation}

Let $\lambda$, $\mu$ and $\nu$ be the Euler angles for the matrix $F$, i.e.,
\[
F = R(\bm n_3, \lambda) R(\bm n_2, \mu) R(\bm n_3, \nu).
\]
{}From Eq.~(\ref{5.21}) it follows that these angles are transformed under
the action of the gauge transformations as follows:
\[
\lambda(\tau) = \lambda, \qquad \mu(\tau) = \mu, \qquad \nu(\tau) = \nu -
S\tau.
\]

It is clear that we can use the quantities $S$, $\varphi$, $\vartheta$ and
$\lambda$, $\mu$, $\nu$ as independent coordinates in the initial phase
space of the system. The symplectic two--form $\omega$, written in terms
of these coordinates has the form
\begin{equation}
\omega = d(S \cos \vartheta d \varphi -  S d\nu - S \cos \mu d\lambda).
\label{5.24}
\end{equation}
To show that Eq.~(\ref{5.24}) is valid we start with expression
(\ref{5.4}). From Eqs.~(\ref{5.19}), (\ref{c.26a}) and (\ref{5.21a}) it
follows that
\[
\omega = \frac{1}{2} d[S {\cal R}(\bm s, \bm n_3)(\bm n_3 \times \bm
f_i)d({\cal R}(\bm s, \bm n_3) \bm f_i)].
\]
Using matrix notation we can rewrite this relation in the form
\begin{equation}
\omega = - \frac{1}{2} d[S\,\mbox{tr}\,(M_3 R^T(\varphi, \vartheta) d
R(\varphi, \vartheta))] + \frac{1}{2} d[S\,\mbox{tr}\,(M_3 F^T(\lambda, \mu,
\nu) dF(\lambda, \mu, \nu))]. \label{d.2}
\end{equation}
Further, Eqs.~(\ref{c.23a}) and (\ref{c.26}) lead to the equalities
\begin{eqnarray*}
&&R^T(\varphi, \vartheta) d R(\varphi, \vartheta) = (-\sin \vartheta M_1 + \cos
\vartheta M_3) d\varphi + M_2 d\vartheta, \label{d.3} \\
&&F^T(\lambda, \mu, \nu) d F(\lambda, \mu, \nu) = [-\sin \mu (cos \nu M_1 -
\sin \nu M_2) + \cos \mu M_3] d\lambda \nonumber \\
&&\hspace{12.0em} + (-\sin \nu M_1 + \cos \nu M_2) d\mu + M_3 d \nu.
\label{d.4}
\end{eqnarray*}
Now using these expressions in Eq.~(\ref{d.2}), and taking into account
Eq.~(\ref{c.23}), we come to Eq.~(\ref{5.24}).

It is instructive to compare Eq.~(\ref{5.24}) with Eq.~(\ref{3.29}). From
Eq.~(\ref{5.24}) we see that the restriction of the symplectic two--form
to the constraint surface, defined by the constraint $\psi$, has the form
\[
\omega = \rho d(\cos \vartheta d \varphi - \cos \mu d \lambda).
\]

Using the definition of the variables $J_i$ and quantities $F_{ij}$,
we see that
\[
J_i = SF_{i3}.
\]
{}From this relation it is easy to conclude that $\lambda$ and $\mu$ can be
considered as the spherical angles, parameterizing $J_i$, that actually are the
polar angles of the vector $\bm S$ with respect to the basis $\{\bm e_i\}$:
\[
\bm S = S(\cos \lambda \sin \mu \bm e_1 + \sin \lambda \sin \mu \bm e_2 +
\cos \mu \bm e_3).
\]
It is quite clear that the quantities $S_i$ and $J_i$ form a set of
dependent coordinates in the reduced phase space under consideration.
Here we have $S_i S_i = J_i J_i = \rho^2$, that confirms the conclusion
made above that the reduced phase space in this case is $S^2 \times S^2$.
Using now the experience gained in the preceding Section, we see that
the symplectic two--form can be written in terms of these coordinates as
\[
\omega = - \frac{1}{2 \rho^2} (\epsilon_{ijk} S_i dS_j \wedge dS_k -
\epsilon_{ijk} J_i dJ_j \wedge dJ_k).
\]

Let us turn our attention to the constraint $\chi$. It is easy to get
convinced that the transformations of the gauge group, generated by this
constraint act in the initial phase space in the following way:
\begin{eqnarray*}
&\bm e_i(\tau) = {\cal R}(\bm e_3, \tau) \bm e_i = \bm e_j R_{ji}(\bm n_3,
\tau),& \\
&\bm S(\tau) = \bm S.&
\end{eqnarray*}
We see that the gauge group, generated by the constraint $\chi$, acts only
in one factor of the product $S^2 \times S^2$, which is a reduced phase
space obtained by us after reduction with the help of the constraint
$\psi$.  Thus we can consider only that factor, which is evidently
described by the quantities $J_i$. From such point of view, the constraint
surface, defined by the constraint $\chi$, is a one dimensional sphere
$S^1$, where the group of gauge transformations acts transitively.  Hence,
after reduction we get only one point. Thus, the final reduced phase space
is a two--dimensional sphere $S^2$.

In matrix notation we have
\[
E(\tau) = R^T(\bm n_3, \tau) E,
\]
that implies
\[
F(\tau) = R^T(\bm n_3, \tau) F.
\]
Hence, for the Euler angles $\lambda$, $\mu$, and $\nu$ we get
\[
\lambda(\tau) = \lambda - \tau, \qquad \mu(\tau) = \mu, \qquad \nu(\tau) =
\nu.
\]
It is clear from this consideration, that the reduced phase space is
described by the independent coordinates $\varphi$, $\vartheta$.
The symplectic two--form on the reduced phase space is
\[
\omega = \rho d(\cos \vartheta d \varphi),
\]
that can be written in terms of the dependent coordinates $S_i$ as
\[
\omega = - \frac{1}{2 \rho^2} \epsilon_{ijk} S_i dS_j \wedge dS_k.
\]

The reduced phase space we have obtained, coincides with the reduced phase
space from the preceding Section. Hence the geometric quantization method
gives again the quantization condition  (\ref{r22}) for the parameter $\rho$,
while the parameter $\kappa$ remains unquantized here. Therefore, while
for this model unlike the previous one, two methods of quantization lead
to the quantum system, describing either integer or half-integer spin
states, nevertheless, the corresponding quantum systems are different: the
Dirac method gives discrete values for the observable $\hat J_3$, whereas
the reduced phase space quantization allows it to take any value $\kappa$,
such that $\kappa^2 < j^2$ for a system with spin $j$.

Let us note here one interesting property of the system.  We can use a
combination of the Dirac and reduced phase space quantization methods.
After the first reduction  with the help of the constraint $\psi$, the system,
described  by the spin vector and the `isospin' vector \cite{ply5} with
the components $I_i = -J_i$, $S_i S_i = I_i I_i$, can be quantized
according to Dirac by imposing the quantum analog of the constraint $\chi$
on the state vectors for singling out the physical states.  In this case
we have again the quantization of the parameter $\kappa$ as in the pure
Dirac quantization method, and, therefore, here the observable $\hat J_3$
can take only integer or half--integer value.  Hence, in this sense, such
a combined method gives the results coinciding with the results of the
Dirac quantization method.

\section{Discussion and Conclusions}

The plane spin model, considered in Section 3, gives an example of the
classical constrained system with finite number of the degrees of freedom
for which there is no gauge condition, but nevertheless, the reduced phase
space can be represented as a submanifold of the constraint surface.  As
we have seen, Dirac and reduced phase space quantization methods lead to
the coinciding physical results for this plane spin model.  Moreover, we
have revealed an interesting analogy in interpretation of the situation
with nonexistence of a global gauge condition for this simple constrained
system with the situation taking place for the non-Abelian gauge theories
\cite{sin}.

The rotator and top spin models give examples of the classical systems, in
which there is no global section of the space of gauge orbits.  In spite
of impossibility to impose gauge conditions such systems admit the
construction of the reduced phase space.  These two models demonstrate
that the reduced phase space and the Dirac quantization methods (or some
their combinations), can give essentially different physical results.

Thus, for Hamiltonian systems with first class constraints we encounter
two related problems.

The first problem consists in the choice of a `correct'
quantization method for such systems. From the mathematical point of view
any quantization leading to a quantum system, which has the initial system
as its classical limit, should be considered as a correct one, but
physical reasonings may distinguish different quantization methods.
Consider, for example, the above mentioned systems. The rotator spin
model, quantized according to the Dirac method, represents by itself the
orbital angular momentum system with additional condition (\ref{r9})
singling out the states with a definite eigenvalue of angular momentum
operator $\hat{\bm S}{}^{2}$. This eigenvalue, in turn, is defined by the
concrete value of the quantized parameter of the model: $\rho^2 =
n(n+1)>0$.  On the other hand, the reduced phase space quantization of the
model gives either integer or half--integer values for the spin of the
system. If we suppose that the system under consideration is to describe
orbital angular momentum, we must take only integer values for the
parameter $\rho$ in the reduced phase space quantization method.
But in this case we must, nevertheless, conclude, that the reduced phase
space quantization method of the rotator spin model describes a more general
system than the quantum system obtained as a result of the Dirac
quantization of that classical system.

The Dirac quantization of the top spin model, or its described combination
with the reduced phase space quantization gives us a possibility to
interpret this system as a system having spin and isospin degrees of
freedom (with equal spin and isospin: $\hat{\bm S}^2=\hat{I}_i\hat{I}_i =
j(j+1)$), but in which the isospin degrees of freedom are
`frozen' by means of the condition $\hat{I}_{3}\Phi_{phys} = -k
\Phi_{phys}$.  On the other hand, as we have seen, the reduced space
quantization method does not allow one to have such interpretation of the
system since it allows the variable $I_{3}$ to take any (continuous) value
$-\kappa$ restricted only by the condition $\kappa^2 < j^2$, i.e.,
the operator $\hat{I}_{3}$ (taking here only one value) cannot be
interpreted as a component of the isospin vector operator. From this point
of view a `more correct' method of quantization is the Dirac quantization
method.

In this respect it is worth to point out that there is
a class of physical models, for which it is impossible to get
the reduced phase space description, and which, therefore, can be
quantized only by the Dirac method.

Indeed, there are various pseudoclassical models containing first class
nilpotent constraints of the form \cite{bri}--\cite{cor}:
\begin{equation}
\psi = \xi_{i_{1}}...\xi_{i_{n}}G^{i_{1}...i_{n}} = 0, \label{7.1}
\end{equation}
where $\xi_{i_{k}}$, are real Grassmann variables with the Poisson brackets
\[
\{\xi_{k},\xi_{l}\}=-ig_{kl},
\]
$g_{kl}$ being a real nondegenerate symmetric constant matrix. Here
it is supposed that
$G^{i_{1}...i_{n}}$, $n \ge 2$, are some functions of other variables,
antisymmetric in their indices, and
all the terms in a sum have simultaneously either even or odd
Grassmann parity.
For our considerations it is important that constraints
(\ref{7.1}) are the constraints, nonlinear in Grassmann variables,
and that they have zero projection on the unit of Grassmann algebra.
In the simplest example of relativistic massless vector particle in
(3+1)--dimensional space--time \cite{bri}--\cite{mar1} the odd part of the
phase space is described by two Grassmann vectors $\xi_{\mu}^{a}$,
$a=1,2$, with brackets
\[
\{\xi_{\mu}^{a},\xi_{\nu}^{b}\}=-i\delta^{ab}g_{\mu\nu},
\]
and the corresponding nilpotent first class constraint has the form:
\begin{equation}
\psi = i\xi_{\mu}^{1}\xi_{\nu}^{2}g^{\mu\nu} = 0, \label{7.4}
\end{equation}
where $g_{\mu\nu}= \mbox{diag}(-1,1,1,1)$.
This constraint is the generator of
the SO(2)--rotations in the `internal isospin' space:
\[
\xi_\mu^1 (\tau) = \xi_\mu^1 \cos \tau + \xi_\mu^2 \sin \tau,
\quad
\xi_\mu^2 (\tau) = \xi_\mu^2 \cos\tau - \xi_\mu^1 \sin \tau.
\]
The specific property of this transformation is
that having $\xi^a_\mu(\tau)$ and $\xi_\mu^a$, we cannot determine the
rotation angle $\tau$ because there is no notion of the inverse element
for an odd Grassmann variable.  Another specific feature of the nilpotent
constraint (\ref{7.4}) is the impossibility to introduce any, even local,
gauge constraint for it. In fact, we cannot find a gauge constraint $\chi$
such that the Poisson bracket $\{\psi,\chi\}$ would be invertible.
Actually, it is impossible {\it in principle} to construct the
corresponding reduced phase space for such a system.
Obviously, the same situation arises for the constraint of general form
(\ref{7.1}).  It is necessary to note here that in the case when the
constraint $\psi$ depends on even variables of the total phase space (see,
e.g., ref. \cite{cor}), and, therefore, generates also transformations of
some of them, we cannot fix the transformation parameter (choose a
point in the orbit) from the transformation law of those even variables,
because the corresponding parameter is present in them with a
noninvertible factor, nonlinear in Grassmann variables.  Therefore, the
pseudoclassical systems containing the constraints of form (\ref{7.1}) can
be quantized only by the Dirac method, that was done in
original papers \cite{spin1,ply1,cor} (see also Ref. \cite{mar1}
where the BRST quantization of such systems was considered).

Let us come back to the discussion of the revealed difference
between two methods of quantization, and point out that
the second related problem is clearing up the sense of gauge degrees of
freedom.
The difference appearing under the Dirac and reduced phase
space quantization methods can be understood as the one proceeding from the
quantum `vacuum' fluctuations corresponding to the `frozen' (gauge)
degrees of freedom.  Though these degrees of freedom are `frozen' by the
first class constraints, they reveal themselves through quantum
fluctuations, and in the Dirac quantization method they cannot be
completely `turned off' due to the quantum uncertainty principle. Thus, we
can suppose that the gauge degrees of freedom serve not simply for
`covariant' description of the system but have `hidden' physical meaning,
in some sense similar to the compactified degrees of freedom in the
Kaluza--Klein theories. If we adopt such a physical point of view, we have
to use only the Dirac quantization method.  Further, the gauge principle
cannot be considered then as a pure technical principle. From here we
arrive also at the conclusion that the Dirac separation of the constraints
into first and second class constraints is not technical, and nature
`distinguish' these two cases as essentially different, since gauge
degrees of freedom, corresponding to the first class constraints, may
reveal themselves at the quantum level (compare with the point of view
advocated in Ref.~\cite{jac}).

Finally, let us mention also that since the path integral quantization
method of constrained systems with first class constraints is based on the
introduction of the gauge conditions for the first class constraints
\cite{Fad69,FPo67}, it appears that the usage of this method for the systems
considered in the present paper is rather problematic.

\vskip 1.em

{\bf Acknowledgements.}
One of the authors (M.P.) thanks J.M. Pons for useful discussions and
kind hospitality during his visit at
Department d'Estructura i Constituents de la Mat\'eria, Universitat de
Barcelona.

\appendix

\section{Pseudoinverse Mappings}

In this appendix and in the next one, we describe the method of dependent
coordinates, that is a very convenient method to deal with second class
constraints (see also \cite{raz}).

Let $A$ be a linear mapping from an $n$--dimensional linear space $V$ to
an $m$--dimensional linear space $W$. It is known that for any
decompositions
\begin{equation}
V = V_0 \oplus V_1, \qquad W = W_0 \oplus W_1, \label{a.1}
\end{equation}
such that $V_0 = \ker A$, $W_1 = \im A$, the mapping $A$ induces an
isomorphism $A'$ of $V_1$ and $W_1$. Here we have
\begin{equation}
A = \iota^{W_1} A' \pi^{V_1}, \label{a.2}
\end{equation}
where $\pi^{V_1}$ is the projection of $V$ onto $V_1$, and $\iota^{W_1}$
is the inclusion mapping of $W_1$ into $W$. Note that decompositions
(\ref{a.1}) generate the dual decompositions
\[
V^* = V^*_0 \oplus V^*_1, \qquad W^* = W^*_0 \oplus W^*_1,
\]
such that
\[
W^*_0 = \ker A^*, \qquad V^*_1 = \im A^*,
\]
where $A^*: W^* \to V^*$ is the adjoint mapping of the mapping $A$.

The pseudoinverse mapping of the mapping $A$ is defined as
\begin{equation}
A^{-1} = \iota^{V_1} A'^{-1} \pi^{W_1}, \label{a.3}
\end{equation}
where $\pi^{W_1}$ is the projection of $W$ onto $W_1$, and $\iota^{V_1}$
is the inclusion mapping of $V_1$ into $V$. From Eqs.~(\ref{a.2}) and
(\ref{a.3}) we get
\[
A^{-1} A = \Pi, \qquad A A^{-1} = \Sigma,
\]
where
\[
\Pi = \iota^{V_1} \pi^{V_1}, \qquad \Sigma = \iota^{W_1} \pi^{W_1}.
\]
For the pseudoinverse mapping $A^{-1}$ we have also
\[
\ker A^{-1} = W_0, \qquad \im A^{-1} = V_1,
\]
furthermore,
\[
\ker A^{-1*} = V^*_0, \qquad \im A^{-1*} = W^*_1.
\]

Choosing different decompositions of the linear spaces $V$ and $W$ we get
different pseudoinverse mappings. Actually, it can be shown that if for
a given mapping $A$ we fix decompositions (\ref{a.1}), then there exists a
unique mapping $A^{-1}$, defined by the conditions
\[
A A^{-1} = \Sigma, \qquad \ker A^{-1*} = V^*_0,
\]
or by the conditions
\[
A^{-1} A = \Pi, \qquad \ker A^{-1} = W_0.
\]

Note that decompositions (\ref{a.1}) can be fixed in the following way.
Denote by $p$ the rank of the mapping $A$. Recall that $\rank A = \dim \im
A$. Let $\{e_\alpha\}_{\alpha = 1,\ldots,n-p}$ be a basis in $V_0 = \ker
A$, and $\{\mu^\alpha\}_{\alpha = 1,\ldots,n-p}$ be a set of elements
of $V^*$ such that
\[
\mu^\alpha(e_\beta) = \delta^\alpha{}_\beta.
\]
In this case we can define $V_1$ as
\[
V_1 = \{ v \in V \mid \mu^\alpha(v) = 0,\, \alpha = 1,\ldots, n-p\},
\]
and we have
\[
\Pi = 1 - e_\alpha \otimes \mu^\alpha.
\]

As $\rank A^* = \rank A$, then $\dim \ker A^* = m-p$. Let
$\{\nu^\sigma\}_{\sigma = 1,\ldots, m-p}$ be a basis in $\ker A^* =
W^*_0$, and $\{f_\sigma\}_{\sigma = 1,\ldots,m-p}$ be a set of
elements of $W$ such that \[ \nu^\tau(f_\sigma) =
\delta^\tau{}_\sigma.  \] The space $W_0$ can be defined as the
space generated by the set $\{f_\sigma\}_{\sigma =
1,\ldots,m-p}$. Indeed, it is not difficult to show that in
this case $W = W_0 \oplus W_1$. The mapping $\Sigma$ in this
case can be presented in the form
\[
\Sigma = 1 - f_\sigma \otimes \nu^\sigma.
\]

Treating an arbitrary $n \by m$ matrix as the matrix of a linear mapping
from an $n$--dimen\-si\-o\-nal linear space to an $m$--dimensional linear
space, we associate  a linear mapping with any matrix. Hence, we can
define the pseudoinverse matrix of a matrix as the matrix of the
corresponding pseudoinverse operator. We can reformulate the results of
the above consideration for the case of matrices as follows.

Let $a = \|a^r{}_i\|$ be $n \by m$ matrix of rank $p$. Hence, there is
a complete set of $n-p$ independent relations of the form
\[
a^r{}_i e^i{}_\alpha = 0, \qquad \alpha = 1,\ldots, n-p.
\]
Since the rank of the transposed matrix equals the rank of the initial
one, there is a complete set of $m-p$ independent relations
\[
\nu^\tau{}_r a^r{}_i = 0, \qquad \tau = 1,\ldots,m-p.
\]
It is clear that there are quantities $\mu^\alpha{}_i$ and $f^r{}_\tau$
such that
\[
\mu^\alpha{}_i e^i{}_\beta = \delta^\alpha{}_\beta, \qquad \nu^\tau{}_r
f^r{}_\sigma = \delta^\tau{}_\sigma.
\]
Define the matrices $\|\pi^i{}_j\|$ and $\|\sigma^r{}_s\|$ by
\[
\pi^i{}_j = \delta^i{}_j - e^i{}_\alpha \mu^\alpha{}_j, \qquad \sigma^r{}_s
= \delta^r{}_s - f^r{}_\tau \nu^\tau{}_s.
\]
{}From the discussion above we see that there exists a unique matrix
$a^{-1} = \|(a^{-1})^i{}_r\|$, defined by the conditions
\[
a^r{}_i (a^{-1})^i{}_s = \sigma^r{}_s, \qquad \mu^\alpha{}_i
(a^{-1})^i{}_r = 0,
\]
or by the conditions
\[
(a^{-1})^i{}_r a^r{}_j = \pi^i{}_j, \qquad (a^{-1})^i{}_r f^r{}_\tau = 0.
\]

Consider a special case of skew symmetric $n \by n$ matrices. Usually,
with such matrices we associate linear operators from an $n$--dimensional
$V$ space to the dual space $V^*$. If a matrix $\|a_{ij}\|$ is skew
symmetric then the corresponding mapping $A$ is skew adjoint, $A^* = -A$.
If we choose in (\ref{a.1})
\[
W_0 = V^*_0, \qquad W_1 = V^*_1,
\]
then the pseudoinverse mapping is skew adjoint and the corresponding
pseudoinverse matrix is skew symmetric.

\section{Dependent Coordinates}

Let $N$ be a smooth manifold of the dimension $n$. Suppose that in some
open neighborhood $U \subset N$ there is given a set of functions $z^i$, $i =
1,\ldots, m > n$ such that for any point $p \in U$ the rank of the set
$dz^i{}_p$, $i = 1, \ldots, m$, is equal to $n$. In such a case we call
the functions $z^i$ local dependent coordinates in $M$. It is clear that
if we consider the mapping $\iota: U \to {\bf R}^m$, defined by
\[
\iota(p) = (z^1(p), \ldots, z^m(p)), \qquad p \in U,
\]
then we can consider $U$ as a smooth $n$--dimensional submanifold of ${\bf
R}^m$. In fact the case when $U = M$, i.e., the case of globally defined
dependent coordinates, is the most interesting one.

Since for any $p \in U$ the rank of the set $dz^i{}_p$ is equal to $n$,
then there exists a set of smooth functions $\mu^\alpha{}_i$, $\alpha = 1,
\ldots, m-n$, such that
\[
\mu^\alpha{}_i dz^i = 0, \qquad \alpha = 1, \ldots, m-n.
\]
It is clear that the quantities $\mu^\alpha{}_i$ can be chosen in such a
way that the set of $m-n$ $m$--dimensional vectors $\mu^\alpha{}_i(p)$
be linearly independent at any point $p\in U$.

Let $u^a$, $a = 1,\ldots, n$, be a set of independent coordinates defined
in some open neighborhood $V \subset U$. For the dependent coordinates
$z^i$ we get
\[
dz^i = z^i{}_a du^a,
\]
where
\[
z^i{}_a = \frac{\partial z^i}{\partial u^a}.
\]
The $m \by n$ matrix $\|z^i{}_a\|$ is of rank $n$, and satisfies the
relations
\[
\mu^\alpha{}_i z^i{}_a = 0.
\]
Let $e^i{}_\alpha$ be a set of smooth functions, such that
\[
\mu^\alpha{}_i e^i{}_\beta = \delta^\alpha{}_\beta.
\]
Define the matrix $\|\pi^i{}_j\|$ by
\[
\pi^i{}_j = \delta^i{}_j - e^i{}_\alpha \mu^\alpha{}_j.
\]
Denote by $z^a{}_i$ the matrix elements of the pseudoinverse matrix of the
matrix $\|z^i{}_a\|$, defined by the relations
\[
z^i{}_a z^a{}_j = \pi^i{}_j.
\]
As it follows from Appendix A, we have the relations
\begin{equation}
z^a{}_i z^i{}_b = \delta^a{}_b, \qquad z^a{}_i e^i{}_\alpha = 0.
\end{equation}

Now we can introduce a set of the vector fields $\partial/\partial z^i$,
defined by
\[
\frac{\partial}{\partial z^i} = z^a{}_i \frac{\partial}{\partial u^a}.
\]
These vector fields satisfy the relations
\[
e^i{}_\alpha \frac{\partial}{\partial z^i} = 0,
\]
and they are, in a sense, dual to the differentials $dz^i$:
\begin{equation}
dz^i\left(\frac{\partial}{\partial z^j}\right) = \pi^i{}_j. \label{b.11}
\end{equation}
It is quite clear that the vector fields $\partial/\partial z^i$ are
independent of the choice of local coordinates $u^a$. An arbitrary vector
field $X$ can be represented as
\[
X = X^i \frac{\partial}{\partial z^i}.
\]
Note that the functions $X^i$ in this relation are defined ambiguously. To
make them unique we subject them to the relations
\begin{equation}
\mu^\alpha{}_i X^i = 0. \label{b.13}
\end{equation}

Analogously, for any one--form $\varphi = \varphi_i dz^i$
the functions $\varphi_i$ are fixed by the relations
\[
\varphi_i e^i{}_\alpha = 0.
\]
Such a fixation allows one to write
\begin{equation}
df = \frac{\partial f}{\partial z^i} dz^i \label{b.16}
\end{equation}
for any function $f$, and
\[
\varphi(X) = \varphi_i X^i
\]
for any one--form $\varphi$ and vector field $X$.

Suppose now that the manifold $N$ is a symplectic manifold with the
symplectic two--form $\omega$, having in the local independent coordinates
$u^a$ the representation
\[
\omega = \frac{1}{2} \omega_{ab} du^a \wedge du^b.
\]
Let $\|\omega^{ij}\|$ be the matrix composed by the Poisson brackets of
the functions $z^i$:
\[
\{z^i, z^j\} = \omega^{ij}.
\]
Using the coordinates $u^a$, we can write
\begin{equation}
\omega^{ij} = \omega^{ab} z^i{}_a z^j{}_b, \label{b.20}
\end{equation}
where $\omega^{ab} = \{u^a, u^b\}$. Hence, we have
\begin{equation}
\mu^\alpha{}_i \omega^{ij} = 0. \label{b.21}
\end{equation}
It is easy to understand that the matrix $\|\omega^{ij}\|$ is of rank $n$,
therefore, the $m$--dimensional vectors $\mu^\alpha{}_i$ form a complete
system of the vectors, satisfying Eq.~(\ref{b.21}).

It is clear that the symplectic two--form $\omega$ can be represented as
\begin{equation}
\omega = \frac{1}{2} \omega_{ij} dz^i \wedge dz^j. \label{b.22}
\end{equation}
The matrix $\|\omega_{ij}\|$ is supposed to be skew symmetric, and is
defined ambiguously. We fix it with the help of the conditions
\begin{equation}
\omega_{ij} e^j{}_\alpha = 0. \label{b.23}
\end{equation}
Writing the obvious relation
\[
\omega_{ij} = \omega_{ab} z^a{}_i z^b{}_j,
\]
and comparing it with Eq.~(\ref{b.20}) we conclude that
\[
\omega^{ik} \omega_{kj} = \pi^i{}_j.
\]
Thus, the matrix $\|\omega_{ij}\|$ is the pseudoinverse matrix of the
matrix $\|\omega^{ij}\|$. Actually, the inverse statement is also true.
Namely, if we have the representation of the symplectic two--form $\omega$
of type (\ref{b.22}), where the skew symmetric matrix
$\|\omega_{ij}\|$ satisfies relations (\ref{b.23}), then the
matrix $\|\omega^{ij}\|$, giving the Poisson brackets of the
functions $z^i$, is the pseudoinverse matrix of the matrix
$\|\omega_{ij}\|$.

The Hamiltonian vector field $X_f$, corresponding to the function $f$ on
$N$, can be written as
\[
X_f = X_f^i \frac{\partial}{\partial z^i}.
\]
Using Eqs.~(\ref{6}), (\ref{b.11}), (\ref{b.13}) and (\ref{b.16}) we get
\[
X_f^i = \frac{\partial f}{\partial z^j} \omega^{ji}.
\]
Hence, we have
\[
\{f, g\} = \frac{\partial f}{\partial z^i} \omega^{ij} \frac{\partial
g}{\partial z^j},
\]
and, in particular,
\begin{equation}
X^i_f = \{f, z^i\}. \label{b.*}
\end{equation}
Thus, our conventions lead to the same formulae that we have in the case
of independent coordinates.

In the present paper we suppose that the one parameter group of canonical
transformations $\Phi_\tau$, generated by a function $\psi$, is defined by
the equation
\[
\dot \Phi_\tau(p) = - X_\psi(\Phi_\tau(p)), \qquad p \in N.
\]
{}From this equation for any function $f$ we have
\begin{equation}
\Phi^*_\tau f = \sum_{k=0}^\infty \frac{\tau^k}{k!}
\underbrace{\{\ldots\{f}_k, \psi\},\ldots, \psi\}. \label{b.25a}
\end{equation}

Consider now the case when the symplectic manifold $N$ is a symplectic
submanifold of an $m$--dimensional symplectic manifold $M$. Let $\Omega$
be the symplectic two--form on $M$, and $\iota: N \to M$ be the inclusion
mapping. In this case we have
\[
\omega = \iota^*\Omega.
\]
We assume further that the submanifold $N$ is a submanifold, defined by the
equations
\[
\psi^\alpha = 0, \qquad \alpha = 1,\ldots,m-n.
\]

For the functions $\psi^\alpha$ we have
\[
\iota^* \psi^\alpha = 0,
\]
then
\begin{equation}
\iota^* d\psi^\alpha = 0. \label{b.29}
\end{equation}
Let $Z^i$, $i=1,\ldots,m$ be a set of independent local coordinates in
$M$. It is clear that the functions $z^i = \iota^* Z^i$ form a set of
dependent coordinates in $N$. Using the representation
\[
d\psi^\alpha = \frac{\partial \psi^\alpha}{\partial Z^i} dZ^i,
\]
from Eq.~(\ref{b.29}) we get
\[
\iota^* \frac{\partial \psi^\alpha}{\partial Z^i} dz^i = 0.
\]
It is clear that we can put
\[
\mu^\alpha{}_i = \iota^* \frac{\partial \psi^\alpha}{\partial Z^i}.
\]

As it was proved in Section 2, at any point $p\in N$ the matrix
\[
\widetilde \Xi^{\alpha\beta}(p) = \iota^* \{\psi^\alpha, \psi^\beta\}(p)
\]
is nondegenerate. Using the coordinates $Z^i$, we can write
\[
\Omega = \frac{1}{2} \Omega_{ij} dZ^i \wedge dZ^j,
\]
hence,
\[
\{Z^i, Z^j\} = \Omega^{ij},
\]
where $\|\Omega^{ij}\|$ is the inverse matrix of the matrix
$\|\Omega_{ij}\|$.  Thus, we have
\[
\widetilde \Xi^{\alpha\beta} = \widetilde \Omega^{ij} \mu^\alpha{}_i
\mu^\beta{}_j, \qquad \widetilde\Omega^{ij} = \iota^* \Omega^{ij}.
\]
This equality allows us to put
\begin{equation}
e^i{}_\alpha = \widetilde \Omega^{ij} \mu^\beta{}_j \widetilde
\Xi_{\beta\alpha}, \label{b.**}
\end{equation}
where $\|\widetilde \Xi_{\alpha\beta}\|$ is the inverse matrix of the matrix
$\|\widetilde \Xi^{\alpha\beta}\|$.

For the symplectic two--form $\omega$ we have
\[
\omega = \widetilde \Omega_{ij} dz^i \wedge dz^j,
\]
where $\widetilde \Omega_{ij} = \iota^* \Omega_{ij}$.
Hence, the functions $\omega_{ij}$, satisfying Eq.~(\ref{b.23}), are given
by
\[
\omega_{ij} = \pi^k{}_i \widetilde \Omega_{kl} \pi^l{}_j = \widetilde
\Omega_{ij} - \mu^\alpha{}_i \widetilde \Xi_{\alpha \beta} \mu^\beta{}_j.
\]
For the pseudoinverse matrix we get
\[
\omega^{ij} = \widetilde \Omega^{ij} - e^i_\alpha \widetilde
\Xi^{\alpha \beta} e^j_\beta,
\]
that with the help of Eq.~(\ref{b.**}) can be written as
\[
\omega^{ij} = \iota^*\left(\Omega^{ij} - \Omega^{ik} \frac{\partial
\psi^\alpha}{\partial Z^k} \Xi_{\alpha \beta} \frac{\partial
\psi^\beta}{\partial Z^l} \Omega^{lj}\right),
\]
or, equivalently,
\[
\{z^i, z^j\} = \iota^* (\{Z^i, Z^j\} - \{Z^i, \psi^\alpha\}
\Xi_{\alpha \beta} \{\psi^\beta, Z^j\}).
\]
{}From this relation for any two functions, $F$ and $G$, on $M$ we get
\[
\{\iota^*F, \iota^*G\} = \iota^*\left(\{F,G\} - \{F, \psi^\alpha\}
\Xi_{\alpha \beta} \{\psi^\beta, G\}\right).
\]
Thus if we define the Dirac bracket of the functions $F$ and $G$ by the
usual relation
\[
\{F, G\}^* = \{F, G\} - \{F, \psi^\alpha\} \Xi_{\alpha
\beta}\{\psi^\beta, G\},
\]
then we have
\[
\{\iota^* F, \iota^* G\} = \iota^*(\{F, G\}^*).
\]

\section{Three--Dimensional Rotation Group}

Consider the space ${\bf R}^3$ as a three--dimensional linear space with
the canonical basis
\[
\bm n_1 = \left(\begin{array}{c}
1 \\ 0 \\ 0
\end{array} \right), \qquad
\bm n_2 = \left(\begin{array}{c}
0 \\ 1 \\ 0
\end{array} \right), \qquad
\bm n_3 = \left(\begin{array}{c}
0 \\ 0 \\ 1
\end{array} \right).
\]
The scalar product of two vectors $\bm u = u_i \bm n_i$ and $\bm v = v_i
\bm n_i$ is defined by
\[
\bm u \bm v = u_1 v_1 + u_2 v_2 + u_3 v_3,
\]
whereas their vector product is
\[
\bm u \times \bm v = \epsilon_{ijk} \bm n_i u_j v_k,
\]
where $\epsilon_{ijk}$ are the components of the totally antisymmetric
tensor with $\epsilon_{123} = +1$.

The three--dimensional rotation group consists of all nondegenerate linear
transformations of ${\bf R}^3$ preserving its orientation and scalar
product of the vectors. This group is denoted SO(3), and its elements are
called rotations.  We specify an orientation in ${\bf R}^3$ considering
the canonical basis as a right one. Note that the canonical basis is
orthonormal. By definition, a rotation $\cal R$ transforms the basis
$\{\bm n_i\}$ to the right orthonormal basis $\{\bm e_i\}$:
\[
\bm e_i = {\cal R} \bm n_i.
\]

We associate the matrix $R$ with the rotation $\cal R$  through
the relation
\begin{equation}
{\cal R} \bm n_i = \bm n_j R_{ji}. \label{c.7}
\end{equation}
As the rotation $\cal R$ preserves the scalar product, the matrix $R$
satisfies the relation
\begin{equation}
R_{ij}R_{kj} = \delta_{ik}, \label{c.8}
\end{equation}
that can be written in matrix notations as
\begin{equation}
R R^T = 1. \label{c.9}
\end{equation}
{}From this relation it follows that
$
\mbox{det }R = \pm 1.
$
Remembering that ${\cal R}$ preserves orientation, we see that actually
\begin{equation}
\mbox{det } R = 1. \label{c.11}
\end{equation}
It is clear that any matrix $R$, satisfying relations (\ref{c.9}) and
(\ref{c.11}), generates a rotation ${\cal R}$, defined by Eq.~(\ref{c.7}).

A simple geometric consideration shows that the rotation ${\cal R}(\bm
n_1, \varphi)$ by the angle $\varphi$ about the direction of $\bm n_1$ is
described by the formulas
\begin{eqnarray*}
{\cal R}(\bm n_1, \varphi) \bm n_1 &=& \bm n_1, \label{c.12} \\
{\cal R}(\bm n_1, \varphi) \bm n_2 &=& \bm n_2 \cos \varphi + \bm n_3 \sin
\varphi, \label{c.13} \\
{\cal R}(\bm n_1, \varphi) \bm n_3 &=& -\bm n_2 \sin \varphi + \bm n_3 \cos
\varphi. \label{c.14}
\end{eqnarray*}
Hence the matrix $R(\bm n_1, \varphi)$ has the form
\[
R(\bm n_1, \varphi) = \left( \begin{array}{ccc}
1 &       0      &      0        \\
0 & \cos \varphi & -\sin \varphi \\
0 & \sin \varphi &  \cos \varphi
\end{array} \right).
\]
Analogously, for $R(\bm n_2, \varphi)$ and $R(\bm n_3, \varphi)$ we get
\[
R(\bm n_2, \varphi) = \left( \begin{array}{ccc}
\cos \varphi  & 0 & \sin \varphi \\
      0       & 1 &       0      \\
-\sin \varphi & 0 & \cos \varphi
\end{array} \right), \qquad
R(\bm n_3, \varphi) = \left( \begin{array}{ccc}
\cos \varphi & -\sin \varphi & 0 \\
\sin \varphi &  \cos \varphi & 0 \\
      0      &        0      & 1
\end{array} \right).
\]

The rotation by the angle $\varphi$ about the direction of a unit vector
$\bm u$ is described by
\begin{equation}
\bm r' = {\cal R}(\bm u, \varphi) \bm r = \bm r \cos \varphi + (\bm u
\times \bm r) \sin \varphi + \bm u (\bm u \bm r)(1 - \cos \varphi).
\label{c.17}
\end{equation}
Hence, we have
\begin{equation}
R_{ij}(\bm u, \varphi) = \delta_{ij} \cos \varphi + \epsilon_{ikj} u_k
\sin \varphi + u_i u_j (1 - \cos \varphi). \label{c.18}
\end{equation}
Let $Q$ be the matrix, corresponding to an arbitrary rotation $\cal Q$. From
(\ref{c.18}) we can easily get
\begin{equation}
Q R(\bm u, \varphi) Q^T = R({\cal Q} \bm u, \varphi). \label{c.18a}
\end{equation}

It is quite clear that the Lie algebra of the group SO(3) is formed by the
matrices $A$, satisfying the relation
\[
A + A^T = 0.
\]
A convenient basis in this Lie algebra is formed by the matrices
\[
M_i = \left. \frac{d R(\bm n_i, \varphi)}{d \varphi} \right|_{\varphi = 0}.
\]
{}From Eq.~(\ref{c.18}) we get
\[
(M_i)_{jk} = \epsilon_{jik} = -\epsilon_{ijk},
\]
and, therefore,
\[
[M_i, M_j] = \epsilon_{ijk} M_k.
\]
A simple calculation gives
\begin{equation}
\mbox{tr }(M_i M_j) = - 2\delta_{ij}. \label{c.23}
\end{equation}
It is clear also that the following equations are valid:
\begin{equation}
\frac{d R(\bm n_i, \varphi)}{d \varphi} = M_i R(\bm n_i, \varphi) = R(\bm
n_i, \varphi) M_i. \label{c.23a}
\end{equation}

Writing relation (\ref{c.11}) in the form
\begin{equation}
\epsilon_{ijk} R_{il} R_{jm} R_{kn} = \epsilon_{lmn}, \label{c.24}
\end{equation}
and using Eq.~(\ref{c.8}), we get
\[
R_{il} \epsilon_{ijk} R_{kn} = R_{jm} \epsilon_{lmn}.
\]
This relation in matrix notation has the form
\begin{equation}
R^T M_i R = R_{ij} M_j. \label{c.26}
\end{equation}
Analogously, from Eq.~(\ref{c.24}) for an arbitrary rotation $\cal R$ and
arbitrary vectors $\bm
u$, $\bm v$ we get the relation
\begin{equation}
({\cal R} \bm u) \times ({\cal R} \bm v) = {\cal R}(\bm u \times \bm v).
\label{c.26a}
\end{equation}

An arbitrary rotation ${\cal R}$ can be parameterized by the Euler angles
with the help of the relation
\[
{\cal R} (\alpha, \beta, \gamma) = {\cal R}(\bm n_3, \alpha) {\cal R}(\bm
n_2, \beta) {\cal R}(\bm n_3, \gamma),
\]
where $0 \le \alpha < 2\pi$, $0 \le \beta \le \pi$, and $0 \le \gamma <
2\pi$.

Let $\{\bm e_i\}$ be an arbitrary orthonormal basis in ${\bf R}^3$. In
this case we have
\begin{equation}
\bm e_i \bm e_j = \delta_{ij}. \label{c.29}
\end{equation}
Expanding the vectors $\bm e_i$ over the canonical basis $\{\bm n_i\}$:
\begin{equation}
\bm e_i = E_{ij} \bm n_j, \label{c.30}
\end{equation}
we get the matrix $E = (E_{ij})$. From Eq.~(\ref{c.29}) it follows that
\begin{equation}
E E^T = 1. \label{c.31}
\end{equation}
The basis $\{\bm e_i\}$ is a right one if and only if
\begin{equation}
\mbox{det }E = 1. \label{c.32}
\end{equation}
Hence, any right orthonormal basis in ${\bf R}^3$ is connected with an
element of the group SO(3). On the other hand, any matrix $E$, satisfying
relations (\ref{c.31}) and (\ref{c.32}), specifies  a right orthonormal
basis in ${\bf R}^3$ by Eq.~(\ref{c.30}). Thus, we see that there is a
one--to--one correspondence between the set of all right orthonormal bases
in ${\bf R}^3$ and the group SO(3).

Note that condition (\ref{c.32}) can be written as
\[
\bm e_i (\bm e_j \times \bm e_k) = \epsilon_{ijk}, \label{c.33}
\]
or as
\[
\bm e_i \times \bm e_j = \epsilon_{ijk} \bm e_k. \label{c.34}
\]

\end{document}